\documentclass{article}
\usepackage{spconf,amsmath,graphicx,multirow}
\usepackage{booktabs}
\usepackage{anyfontsize}

\usepackage{multirow}
\usepackage{caption}
\usepackage{subfigure}
\usepackage{algorithm}
\usepackage{algorithmic}
\usepackage{url}
\usepackage{graphicx}
\usepackage{arydshln}

\usepackage{booktabs}

\ninept
\title{Graph Convolutional Network Based Semi-supervised Learning on Multi-speaker Meeting Data}
\name{Fuchuan Tong$^{\dag 1}$, Siqi Zheng$^2$, Min Zhang$^4$, Yafeng Chen$^{\dag 2}$, Hongbin Suo$^2$, Qingyang Hong$^{3}$, Lin Li$^{1}$
\thanks{\dag  The work was done during an internship at Alibaba.}
\thanks{
This research is funded by the National Natural Science Foundation of China (Grant
No. 61876160 and No. 62001405) and Fundamental Research Funds for the Central Universities (No. 20720210087).
}}
\address{
  $^1$School of Electronic Science and Engineering, Xiamen University, China \\
  $^2$Speech Lab, Alibaba Group, $^3$School of Informatics, Xiamen University, China \\
  $^4$College of Information Science and Electronic Engineering, Zhejiang University, China \\
  \{qyhong, lilin\}@xmu.edu.cn
  }

\email{\{qyhong, lilin\}@xmu.edu.cn}

\begin{document}

\maketitle
\begin{abstract}
  Unsupervised clustering on speakers is becoming increasingly important for its potential uses in semi-supervised learning. In reality, we are often presented with enormous amounts of unlabeled data from multi-party meetings and discussions. An effective unsupervised clustering approach would allow us to significantly increase the amount of training data without additional costs for annotations.
  Recently, methods based on graph convolutional networks (GCN) have received growing attention for unsupervised clustering,  as these methods exploit the connectivity patterns between nodes to improve learning performance. In this work, we present a GCN-based approach for semi-supervised learning.  Given a pre-trained embedding extractor, a graph convolutional network is trained on the labeled data and clusters  unlabeled data  with ``pseudo-labels''. We present a self-correcting training mechanism that iteratively runs the cluster-train-correct process on pseudo-labels.  We show that this proposed approach effectively uses unlabeled data and improves speaker recognition accuracy.
\end{abstract}

\section{INTRODUCTION}

The availability of large labeled datasets and data augmentation methods have spurred remarkable improvements in speaker verification \cite{snyder2018x} \cite{villalba2020state}\cite{wang2020data} \cite{huang2021synth2aug}. However, collecting large amounts of labeled data in the real world is laborious and expensive.
Semi-supervised learning (SSL) \cite{zhu2009introduction} is a technique to utilize unlabeled datasets, mitigating the reliance on annotations. Semi-supervised learning utilizes limited labeled data in tandem with abundant unlabeled data to obtain higher performances. Stronger performance has been observed when unlabeled data is incorporated \cite{zheng2019autoencoder} \cite{zhang2020neighborhood} \cite{zheng2019towards}.

In recent years, we have witnessed the rising popularity of graph convolutional networks in clustering and semi-supervised learning.
By constructing sub-graphs around each instance and its neighbors, Wang et al. \cite{wang2019linkage} proposed formulating the clustering problem as a linkage prediction problem and utilized a graph convolutional network to predict  links between two unlabeled samples. At the same time, Li et al. \cite{yang2019learning} proposed a graph convolutional network that combines  a detection and segmentation pipeline  to cluster samples instead of relying on hand-crafted criteria. Similarly, a few graph-based SSL methods  have also been proposed in the field of speech processing. Yuzong et al. \cite{liu2013graph} demonstrated the power of graph-based SSL systems to outperform standard baseline classifiers by improving phone and segment classification. Besides, GCN-based learning (GBL) algorithms \cite{alexandrescu2007graph} \cite{kirchhoff2011phonetic} have shown to improve phonetic classification performances even more than supervised algorithms. In 2021, Chen et al. \cite{chen2021graph} proposed a graph-SSL method based on label propagation for speaker identification, inferring labels by leveraging unlabeled data. In this work, we investigate GCN-based approaches for clustering speaker embeddings. Specifically, we aim to improve clustering accuracy on multi-speaker meeting data using GCN. This allows us to acquire speaker identifications for every segment in multi-party meetings, hence making use of tremendous amounts of unlabeled data for semi-supervised training.

Existing clustering methods often resort to unsupervised methods, such as K-means \cite{lloyd1982least}, Spectral Clustering \cite{shi2000normalized}, and Agglomerative Hierarchical Clustering (AHC) \cite{zhou2016method}. Spectral Clustering and AHC are currently two common clustering methods in speaker diarization \cite{sell2016priors} \cite{han2008strategies} \cite{xiao2021microsoft} \cite{lin2020self} \cite{lin2019lstm}.
However, K-means implicitly assumes that the samples in each cluster are uniformly distributed around the centroid, and the cluster has to be convex-shaped; Spectral Clustering requires the cluster sizes to be relatively balanced \cite{yang2019learning}. These underlying assumptions often fail in deep speaker embeddings. Consequently, these conventional clustering methods lack the capability to cope with complicated clustered structures, thus often giving rise to noisy clusters, especially when applied to large-scale datasets collected from real-world settings. This problem limits the clustering performance.
 \begin{figure*}[t]
  \centering
\includegraphics[width=0.6\textwidth]{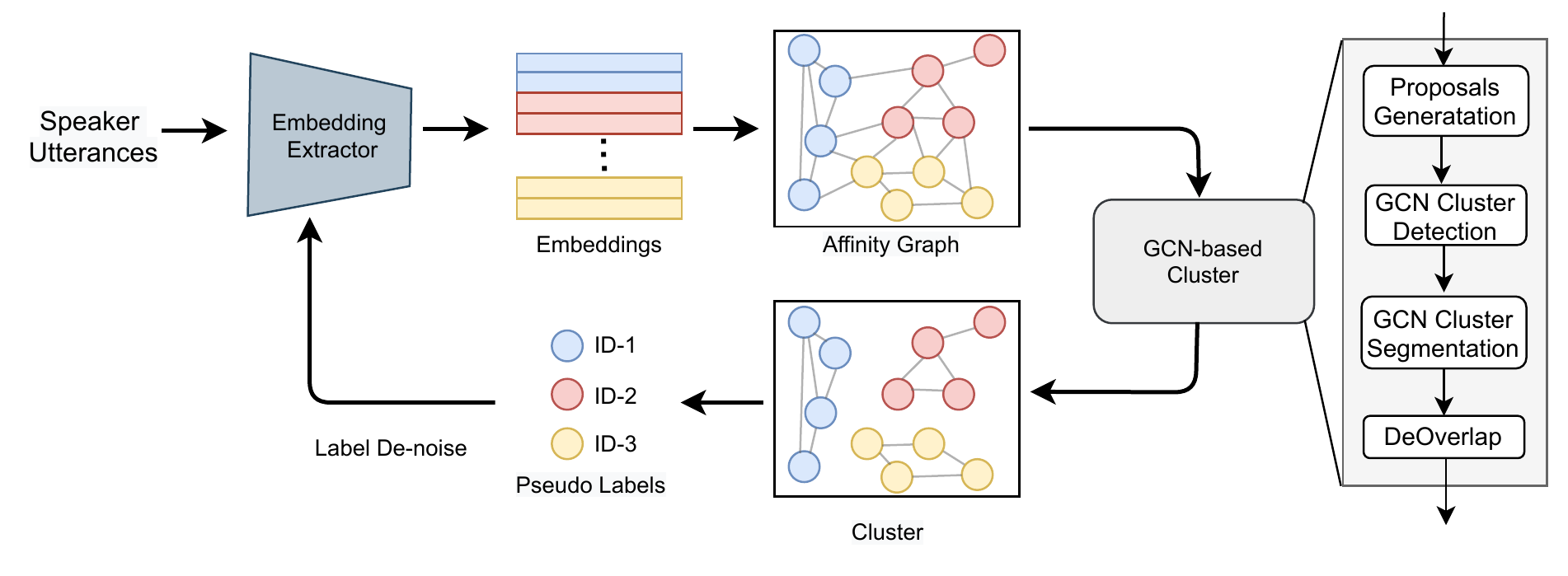}
\caption{The pipeline of our structure. Utterances are first fed into feature extractors to obtain speaker embeddings. An affinity graph is constructed to perform clustering. The cluster results  with pseudo-labels are applied to re-training the deep embedding extractor.}
\label{fig:network}
\end{figure*}
Therefore, to effectively exploit unlabeled speaker data, we need an effective clustering algorithm that is able to cope with the complicated clustered structures that frequently arise in practical scenarios.

 In this work, we present techniques for exploiting unlabeled meeting data. We demonstrate that learning from unlabeled utterances is indeed a
 practical avenue for improving speaker verification. We ameliorate the effect of noisy labels when training on pseudo-labeled data by resorting to a graph convolutional neural network.  To deal with the label noise from the clustering algorithm, we correct  the label noise during the re-training loop by introducing network predictions into a loss function to reduce the effects of erroneous gradients caused by noisy labeled data.

 This paper is organized as follows. In Section 2, we revisit graph convolutional networks. The system  and method
are described in Section 3, and experimental results are presented in Section 4. Finally, Section 5 concludes this work.


\section{REVIEW of GRAPH CONVOLUTIONAL NETWORK}
The recently proposed graph convolutional networks (GCN) \cite{kipf2016semi} \cite{gao2018large} algorithm has demonstrated superior performance in semi-supervised learning on
graph-structured data. GCN is proposed to tackle problems with non-Euclidean data and has a strong capability for modeling graph-structured data; thus, it is quite suitable for label estimation of unlabeled samples. Generally, the basic idea of GCN is to use deep neural networks to map linear and nonlinear relationships in data representations. In each layer, data features are propagated using  graph edges. Specifically,
a propagation can be divided into the following three steps:

(1) Normalization: the input adjacency matrix $W$ is normalized by adding a self-loop to each vertex, resulting in a new adjacency matrix $\bar{W}=W+I$ with the degree matrix $\bar{D}=D+I$, where $I$ is the unit matrix.

(2) Propagation: the layer-wise propagation has the following form:
\begin{equation}
  H^{(k+1)}=\sigma \left( \bar{W}_sH^{(k)}\Theta ^{(k)} \right)
\end{equation}
where $H^{(k+1)}\in R^{N \times d}$  denotes the input data with $N$ samples and $d$ dimensions in the $(k+1)$ layer,
$ \bar{W}_s=\bar{D}^{-\frac{1}{2}}\bar{W}\bar{D}^{-\frac{1}{2}}$ denotes the renormalized graph matrix associated with the data,
 $\Theta^{(k)}$ is the trainable weight matrix in the layer, and $\sigma$ is the activation function.

 (3) Prediction: this step stacks two layers and applies a softmax function on the output features to produce a prediction matrix:
 \begin{equation}
  F=\operatorname{softmax}\left(\tilde{W}_{s}\sigma\left(\tilde{W}_{s} H^{(0)} \Theta^{(0)}\right) \Theta^{(1)}\right)
\end{equation}
 In semi-supervised learning, $F\in R^{C \times N}$ can be seen as the matrix of labels, representing the unknown output of the GCN. The GCN is  learned by forcing  estimated labels to be as close as possible to their expected values.  This is achieved by minimizing the cross-entropy loss in labeled samples.

\section{GCN-BASED SEMI-SUPERVISED LEARNING}
\subsection{GCN-based clustering}
In order to utilize a large amount of unlabeled meeting data for speaker recognition network training, a natural idea is to cluster the speakers from each meeting into pseudo-classes and then treat them  as labeled data to train the speaker network. To  formulate this, suppose that  the labeled speaker data is formulated as $\mathcal{X}^{\mathcal{L}}\in\left\{\mathbf{x}_{1}, \ldots, \mathbf{x}_{l}\right\}$ with sizes $L$, and $\mathcal{Y}^{\mathcal{L}}\in\left\{\mathbf{y}_{1}, \ldots, \mathbf{y}_{l}\right\}$ representing the identity labels. Meanwhile,  the unlabeled meeting data of size $U$ is formulated as $\mathcal{X}^{\mathcal{U}}\in\left\{\mathbf{x}_{l+1}, \ldots, \mathbf{x}_{l+u}\right\}$. The problem is  predicting the labels $\mathcal{Y}^{\mathcal{U}}\in\left\{\mathbf{y}_{l+1}, \ldots, \mathbf{y}_{l+u}\right\}$ since the existing conventional clustering methods lack the capability of coping with complicated clustered structures. In order to solve this problem, we take inspiration from the strengths of graph convolutional networks to deal with graph-structured data. We use the recently proposed supervised graph convolutional networks \cite{yang2019learning} to cluster speaker embeddings of $\mathcal{X}^{\mathcal{U}}$. The overall block diagram of our system is shown in Figure 1. Firstly, we train a speaker embedding extractor on $(\mathcal{X}^{\mathcal{L}},\mathcal{Y}^{\mathcal{L}}) $ in a  supervised fashion. Then, the speaker utterances $\left[ \mathcal{X}^{\mathcal{L}},\mathcal{X}^{\mathcal{U}}\right] $ are fed into an extractor network, and the speaker embeddings are obtained. The embeddings of $\mathcal{X}^{\mathcal{L}}$ and their corresponding labels are used to train a  graph convolutional network. After that,
the affinity matrix of $\mathcal{X}^{\mathcal{U}}$ is  constructed to perform clustering. The clustered results are used to assign pseudo-labels $\mathcal{Y}^{\mathcal{U}}$ for $\mathcal{X}^{\mathcal{U}}$. In  semi-supervised learning processing, the speaker recognition model is re-trained on  all utterances  $\left[ \mathcal{X}^{\mathcal{L}}, \mathcal{X}^{\mathcal{U}}\right] $.

Here, we introduce the overview of the GCN-based clustering \cite{yang2019learning} used in the pipeline. The training procedure consists of the following steps.

\textbf{Affinity Graph Construction.} The affinity graph $\mathcal{G}=(\mathcal{V}, \mathcal{E})$ is  a graph where the nodes $\mathcal{V}$ represent samples,  and the edges $\mathcal{E}$ represent the similarity between the data. Based on the embeddings extracted from a pre-trained speaker embedding extractor model, each sample is regarded as a vertex and  the cosine similarity is used to find $K$ nearest neighbors for each sample. By connecting the neighbors, an affinity graph  is constructed encompassing all samples.

\textbf{Cluster Proposal Generation.} A cluster proposal $\mathcal{P}_{i}$ is a sub-graph of the affinity graph, and it is generated based on super-vertices. A super-vertex contains a small number of vertices that are closely connected to each other. By setting various thresholds on the edge weights of this graph, which can be seen as a preliminary clustering step, a set of super-vertices is generated. Although the samples in the same super-vertex are possible for the  same person, each speaker may contain more than one super-vertex; therefore, further clustering of the super-vertices is needed. Thus, a higher level graph based on a super-vertex is constructed, which takes the centroid of a super-vertex as the vertex and the edges between the centroids as the edges.

\textbf{Cluster Detection.}
This step  determines the likelihood of whether a proposal is correct.  A graph convolutional network is used to extract features for each proposal. The computation of each GCN layer can be formulated as:
\begin{equation}
H^{(k+1)}\left(\mathcal{P}_{i}\right)=\sigma\left(\tilde{D}\left(\mathcal{P}_{i}\right)^{-1}\left(W\left(\mathcal{P}_{i}\right)+I\right) H^{(k)}\left(\mathcal{P}_{i}\right) \Theta^{(k)}\right)
\end{equation}
where $\tilde{D}=\sum_{j} \tilde{W}_{i j}\left(\mathcal{P}_{i}\right)$ is a diagonal degree matrix. $H^{(k)}\left(\mathcal{P}_{i}\right)$ contains the feature vectors in the $k$-th layer. $\Theta^{(k)}$ is a matrix to transform the vectors, and $\sigma$ is the ReLU nonlinear activation function. During training, the GCN is optimized by minimizing an objective function of  the mean square error (MSE) between ground-truth and predicted scores. Then,  high-quality clusters are selected from the generated cluster proposals.

\textbf{Cluster Segmentation.}
Even if the high-quality clusters  are identified by GCN, they  may not be completely pure and could  still contain a few outliers, which must be eliminated. Another similar GCN is developed to  exclude  outliers from the proposal.  In the process of cluster detection, each sample from the retained proposals consists of a set of feature vectors (representing a vertex), an affinity matrix, and a binary vector, which indicates whether the vertex is positive or not. Then, the cluster segmentation model is trained using the vertex-wise binary cross-entropy as the loss function. The segmentation model outputs a probability value for each vertex to indicate how likely it is to be a genuine member instead of an outlier. In the model predictions,  the outliers are removed from the proposals.

\textbf{De-Overlapping.}
There should be no overlapping between the different clusters, so during testing,  a ``de-overlap" procedure uses the predicted GCN scores and sorts them in descending order. The highest GCN score is selected for the proposals to partition the unlabeled dataset into a proper cluster.

The above steps provide a brief overview of the approach for GCN-based clustering. The method learns the structural features of the graph for direct clustering, without assuming that the data obeys a specific distribution; therefore, this method is  beneficial for handling complicated speaker embeddings.

\subsection{Learning with label noise}

Label noise has a significant effect on the performance of speaker embeddings obtained from speaker recognition models
trained on large datasets; this effect has been extensively studied in our previous work \cite{tong21_interspeech}.  The pseudo-labels assigned by clustering inevitably introduce noisy labels  and are quickly memorized by a deep speaker embedding extractor network. To overcome this issue, we utilize our previously proposed approach to learning with label noise. We provide a brief overview of the approach below.
 To avoid a network fitting into noisy samples, the parameters of a network are updated by an improved loss function, which is formulated as:
 \begin{equation}
  L=-\frac{1}{B}\sum_{i=1}^B{\left\{ \left( 1-\alpha_t \right) \log \left( P_{i,y_i} \right) +\alpha_t  \log \left( P_{i,\hat{y}_i} \right) \right\}}
  \end{equation}
where $B$ is the mini-batch size, $P_{i,y_i}$  denotes the posterior probability of sample $\mathbf{x}_i$ being classified as the ground-truth label $y_i$, and $P_{i,\hat{y}_i}$ is a posterior probability of sample $x_i$ being classified as the prediction label. $ \hat{y}_i $ denotes the prediction label for $x_i$. $\alpha_t \in[0,1]$ is the $t$-th training iteration confidence weight  between $P_{i,y_i}$ and $P_{i,\hat{y}_i}$, which determines whether the loss function relies more on the ground-truth label or the predicted label. Specifically, $\alpha_t$ is dynamically increased  and formulated as:

 \begin{equation}
 \alpha_{t}= \alpha_{T}\cdot (t / T)^{\lambda}
    \end{equation}
    where $\alpha_{T}\in[0,1]$ represents the value of $\alpha_t$ at the final iteration. The dynamically increased confident weight means the loss tends to rely more on predicted labels since the  predictions become more and more accurate. During training, sample labels are learned on the fly  to  prevent a network from overfitting incorrect samples. We call this method \textit{ label de-noising} in this paper, and the readers can see \cite{tong21_interspeech} for more details.
\section{EXPERIMENTS}

\subsection{Datasets and experimental setup}

\begin{table}[htbp]
  \setlength{\abovecaptionskip}{0.cm}
      \centering
      \caption{Statistics of the VoxCeleb datasets}
      \resizebox{\linewidth}{!}{
        \begin{tabular}{lccc}
        \hline
        Dateset & \multicolumn{1}{l}{Voxceleb1 dev} & \multicolumn{1}{l}{Voxceleb2 dev} & \multicolumn{1}{l}{Voxceleb1 test} \\
        \hline
        \# of speakers & 1,211  & 5,994  & 40 \\
        \# of utterances & 148,801 & 1,092,009 & 4,715 \\
        Avg. \# of utterances per speaker & 122   & 182   & 112 \\
        Avg. length of utterances & 8 s   & 8 s   & 8 s \\
        \hline
        \end{tabular}%
      \label{tab:addlabel}%
      }
    \end{table}%

We simulate reality using the publicly-available VoxCeleb dataset to ensure reproducibility. In practice, we are often equipped with a small amount of labeled data, represented by Voxceleb1, and large amounts of unlabeled meeting data, simulated by Voxceleb2. The details of the datasets are listed in Table 1.

We show the effectiveness of the GCN-based semi-supervised learning in terms of \textit{speaker clustering}, and \textit{speaker recognition}.
For speaker clustering, we
train the speaker network on a labeled VoxCeleb1 development set to extract speaker embedding. Then, we  use the extracted features  to construct a graph convolutional network and  evaluate speaker clustering performance  with different clustering methods on the VoxCeleb1 test set.
For semi-supervised speaker recognition learning, we use the VoxCeleb2 dataset to simulate unlabeled data from different meetings and perform semi-supervised learning by assigning additional pseudo-labels using various clustering methods.

Data augmentation is also a common approach for augmenting training data in speaker embedding models. Therefore, we adopt similar strategies in \cite{snyder2018x} to augment data, including simulating reverberation with the RIR dataset \cite{ko2017study} and adding noise with the MUSAN dataset \cite{snyder2015musan}. During  training, we use 40-dimensional Mel-frequency cepstral coefficients for each utterance. We  apply the extended time delay network (E-TDNN) described in \cite{snyder2019speaker} to extract embeddings  and choose the AM-softmax loss function to classify speakers.

The codes of the GCN-based clustering method are available on website\footnote{https://github.com/yl-1993/learn-to-cluster/tree/master/dsgcn}. This method is compared to three of the most commonly used clustering algorithms: K-means, Spectral Clustering, and AHC. To measure the clustering performance, we take the evaluation metrics in terms of the pairwise precision, pairwise recall, and F-score, which takes precision and recall into account. The speaker similarity is measured by cosine scoring, and the performance is reported in terms of Equal Error Rate (EER) and detection cost
function (minDCF, set $P_{target}=0.01$).


\subsection{Speaker clustering}
After a network is trained, the clustering algorithms are evaluated in a simulated meeting scenario. To simulate meeting data, we randomly select three test groups from the VoxCeleb1 test set with different number of speakers. Considering that the number of speakers in a real office meeting is usually less than ten, we set the numbers of speakers in each group  to three, six, and nine. To reduce random results during the experiments, we run ten tests on each group and evaluate the performance in terms of  the average precision, recall, and F-score. For all methods, we tune the hyperparameters and report the best results. For K-means and Spectral Clustering (SC), the performance is greatly influenced by the number of  clusters $k$; we vary $k$ for different clusters and report the best results with a high F-score.
The performance of different clustering methods is compared in Table 2.

As shown in Table 2, K-means is highly precise but has poor recall and so yields  an unsatisfactory F-score. This may be due to the assumption  that the samples in each cluster huddle around a single-center in the K-means method. In reality, the speaker embeddings do not satisfy this distribution. Similarly, Spectral Clustering requires clustered data to be balanced, so it is not as precise  when dealing with real-world  data. While AHC is based on a clustering rule that  merges close clusters, it does not require the data  to satisfy certain distribution assumptions so it is  better in terms of precision, recall, and F-score. Similarly, since there are no data distribution assumptions in the GCN-based clustering method, and the clustering rules are learned from the data, the GCN-based clustering method outperforms the AHC. Results in Table 2 demonstrate the effectiveness of GCN-based clustering in capturing complex distributions and improving the precision and recall simultaneously.
\begin{table}[htbp]
  \setlength{\abovecaptionskip}{0.cm}
  \centering
  \caption{Comparison of speaker clustering when the number of clusters is 3, 6, and 9. The results  are the average of the clustering results on 10 different sets of testing data.}
    \begin{tabular}{clccc}
    \hline
    \# & Methods & Precision & Recall & F-score \\
    \hline
    \hline

    \multirow{4}[1]{*}{3} & K-means & 0.80  & 0.52  & 0.63  \\
          & SC & 0.76  & 0.68  & 0.71  \\
          & AHC   & 0.75  & 0.77  & 0.75  \\
          & GCN   & 0.82  & 0.79  & 0.80  \\
    \hline
    \multirow{4}[2]{*}{6} & K-means & 0.78  & 0.56  & 0.65  \\
          & SC & 0.71  & 0.65  & 0.67  \\
          & AHC   & 0.77  & 0.79  & 0.78  \\
          & GCN   & 0.84  & 0.78  & 0.81  \\
    \hline
    \multirow{4}[2]{*}{9} & K-means & 0.77  & 0.53  & 0.63  \\
          & SC & 0.73  & 0.66  & 0.69  \\
          & AHC   & 0.82  & 0.76  & 0.78  \\
          & GCN   & 0.85  & 0.80  & 0.82  \\

    \hline
    \end{tabular}%
  \label{tab:addlabel}%
\end{table}%

\subsection{Semi-supervised speaker recognition}
We assume that there is a small amount of labeled training data and a large amount of unlabeled meeting data. Based on the experiments on speaker clustering in Section 4.2, here we show how to leverage unlabeled data to further boost performance. Firstly, the 5,994 speakers in the VoxCeleb2 development set are randomly shuffled and sampled into 666 meetings without overlapped identities. The number of speakers in the meeting data ranges from two to ten. Then, we utilize the VoxCeleb1 training dataset as in the previous section. After training the initial model with labeled data, we extract speaker embeddings from segments of unlabeled meeting data and cluster them with GCN. Then, we fine-tune the network using  VoxCeleb1 along with the assigned pseudo-labels datasets.

The average precision,  recall, and F-score for applications of different clustering methods on the meeting data are shown in Table \ref{tab:addlabe3}. We also report the  EER and minDCF results of semi-supervised  speaker learning on the VoxCeleb1 test set.
The baseline is  evaluated by a model  with fully-supervised learning and limited labeled VoxCeleb1 training data.
The upper bound is the performance of the model trained with the entirely labeled Voxceleb1 and VoxCeleb2 datasets.
 We observe in the middle part of Table \ref{tab:addlabe3} that the performance of the baseline model is boosted by conducting pseudo-label assignments for unlabeled datasets and semi-supervised training. The results also  show that the higher the F-score of the clusters, the better the EER and minDCF of the model. Due to the optimal clustering performance of the GCN-based method, the performance of the speaker recognition model is  significantly boosted; the EER of the baseline model decreases relatively by 46.6\%, and the minDCF decreases by 50.6\%. However,  there is a slight gap between the results and the model (Oracle) of fully supervised training. This  may be due to the fact that the clustering algorithm inevitably introduces some erroneous samples in certain classes, which is the prime cause for performance deterioration in speaker recognition.  It can be observed that adopting label de-noising (denoted as GCN* in Table \ref{tab:addlabe3})  helps deal with clustering label noise and alleviates the adverse effects. Label de-noising also achieves comparable performance when compared to fully supervised learning.

\begin{table}[htbp]
  \setlength{\abovecaptionskip}{0.cm}
      \centering
      \caption{Performance comparisons of clustering and speaker recognition results using models trained with different clustering pseudo-labels. The * symbol indicates that label de-noising was employed.}
      \resizebox{\linewidth}{!}{\begin{tabular}{lccccc}
        \hline
        Model & \multicolumn{1}{l}{Precision} & \multicolumn{1}{l}{Recall} & \multicolumn{1}{l}{F-score} & \multicolumn{1}{l}{EER} & \multicolumn{1}{l}{minDCF} \\
        \hline
        \hline
        Baseline & -     & -     & -     & 3.34  & 0.384 \\
        \hline
        + \ K-means & 0.78  & 0.54  & 0.64  & 2.04  & 0.255 \\
        + \ SC & 0.74  & 0.67  & 0.70  & 1.73  & 0.213 \\
        + \ AHC  & 0.79  & 0.77  & 0.77  & 1.51  & 0.186 \\
        + \ GCN  & 0.83  & 0.79  & 0.81  & 1.43  & 0.174 \\

        + \ GCN*  & 0.83  & 0.79  & 0.81  & 1.30  & 0.152 \\
        \hline
        Oracle  & -     & -     & -   & 1.28  & 0.165 \\
        \hline
        \end{tabular}%
      \label{tab:addlabe3}%
      }
    \end{table}%

\section{CONCLUSIONS}

We provide a novel approach to improve speaker recognition by leveraging large amounts of unlabeled data. By applying graph convolutional networks on constructed affinity graphs, representations of speaker embeddings are derived by fully utilizing local sub-graph information. Pseudo-labels are generated from cluster predictions on unlabeled data.
Experimental results show that GCN-based clustering  outperforms the existing clustering methods, and the results demonstrate its effectiveness in semi-supervised learning. When combining the clustering method with label de-noising processing, this system achieves  comparable results compared to  fully-supervised training on the Voxceleb1 and Voxceleb2 datasets.  We conclude that the GCN-based clustering method is an  effective method to provide insights into the practice of speaker recognition with unlabeled data.



\bibliographystyle{IEEEtran}

\bibliography{mybib}

\end{document}